# *t*-multiple discrete logarithm problem and solving difficulty

Xiangqun Fu [1,2,*], Wansu Bao [1,2], Jianhong Shi [1,2], Xiang Wang [1,2]
[1] Information Engineering University, Zhengzhou, 450004, China
[2] Synergetic Innovation Center of Quantum Information and Quantum Physics, University of Science and Technology of China, Hefei, Anhui 230026, China

**Abstract**: Considering the difficult problem under classical computing model can be solved by the quantum algorithm in polynomial time, *t*-multiple discrete logarithm problem is presented. The problem is non-degeneracy and unique solution. We talk about what the parameter effects the problem solving difficulty. Then we pointed out that the index-calculus algorithm is not suitable for the problem, and two sufficient conditions of resisting to the quantum algorithm for the hidden subgroup problem are given.
**Keywords**: quantum algorithm, discrete logarithm problem, $t$-multiple discrete logarithm problem, hidden subgroup problem, cryptography

## 1 Introduction

Quantum computation is an entirely new mode. Deutsch algorithm is the first quantum algorithm [1], which shows the power of parallelism computation. And the research on the quantum computation has been attracted widespread attention. Shor's quantum algorithm [2] and Grover's quantum search algorithm [3] have been presented, especially Shor's algorithm which can solve integer factorization and discrete logarithm problem in polynomial time. Then the public-key crypto is under serious threat. Shor's algorithm can be reduced to the quantum algorithm for the hidden subgroup problem [4], which is a pervasive quantum algorithm with polynomial time. Thus the problem can be solved in polynomial time, which can be reduced to hidden subgroup problem. Subsequently, more and more quantum algorithms have been presented [5]~[8]. However, whether is there an effective quantum algorithm for the difficult problem under classical computing mode? NPC problem (Non-deterministic Polynomial-complete problem) can't be efficiently solved on quantum computer. Thus NPC problem is the preferred security basis of the cryptographic algorithm, which can resist the quantum computing attack.

At present, public-key crypto based on NPC problem mainly has four categories: public-key crypto based on error correction of coding, braid group, multiple variant equation and lattice. The public-key crypto based on error correction of coding has more secret keys, which is not practical. The security of the public-key crypto based on braid group and multiple variant equation are be called into question. And if the public-key crypto based on lattice with special properties, it is vulnerable, such as AD public-key crypto [9]. Thus finding a new difficult problem is worth further studying.

In this paper, $t$-multiple discrete logarithm problem ($t$-MDLP) is presented. The solving difficulty of *t*-MDLP is analyzed. If the parameters don't satisfy two sufficient conditions, the problem can't resist the quantum algorithm for the hidden subgroup problem.

## 2 Backgrounds

Discrete logarithm problem is difficult under classical computing model.

***Definition 1**(Discrete logarithm problem)*[10] *Let $G$ be a cyclic group of order $n$. And $\alpha$ is a generator of $G$. The discrete logarithm of $\gamma \in G$ to the base $\alpha$, denoted $\log_\alpha \gamma$, is the unique integer $x$, $0 \le x \le n-1$, such that $\alpha^x = \gamma$.*

At present, index-calculus algorithm [10] is the optimal algorithm for the discrete logarithm problem, which is as follows.

A factor base $S = \{p_1, p_2, \cdots, p_m\}$ is the subset of the group $G$. And a "significant proportion" of all elements in $G$ can be efficiently expressed as a product of elements from $S$.

For arbitrary $k$, $\alpha^k$ is tried to write as a product of elements in $S$:

$$\alpha^k = p_1^{a_1} p_2^{a_2} \cdots p_m^{a_m}, \quad a_i \ge 0.$$

Furthermore, we can obtain a linear relation

$$k = \sum_{i=1}^{m} a_i \log_\alpha p_i \pmod{n}. \tag{1}$$

And $m + c$ relations of the form (1) can be obtained ($c$ is a small positive integer, e.g. $c = 10$, such that the

---

system of equations given by the $m+c$ relations has a unique solution $\log_\alpha p_1(\bmod n), \log_\alpha p_2(\bmod n), \cdots, \log_\alpha p_m(\bmod n)$ with high probability).

Based on the factor base, the discrete logarithm problem can be solved. For arbitrary $\delta$, $\gamma\alpha^\delta$ is tried to write as a product of elements in $S$:

$$\gamma\alpha^\delta = p_1^{b_1} p_2^{b_2} \cdots p_m^{b_m}.$$

Then, we can obtain

$$\log_\alpha \gamma = (-\delta + \sum b_i \log_\alpha p_i)(\bmod n).$$

Thus, the index-calculus algorithm is shown as follows.

*Step*1 Choose a factor base $S = \{p_1, p_2, \cdots, p_m\}$.

*Step*2 Construct the linear relation of $\log_\alpha p_1(\bmod n), \log_\alpha p_2(\bmod n), \cdots, \log_\alpha p_m(\bmod n)$.

*Step*3 Repeat *Step*2, obtain the linear systems of $m+c$ equations. Then a unique solution $\log_\alpha p_1(\bmod n), \log_\alpha p_2(\bmod n), \cdots, \log_\alpha p_m(\bmod n)$ can be obtained with high probability.

*Step*4 For arbitrary $\delta$, compute $\gamma\alpha^\delta$, which can be tried to write as a product of elements in $S$:

$$\gamma\alpha^\delta = p_1^{b_1} p_2^{b_2} \cdots p_m^{b_m}.$$

*Step*5 According to $\log_\alpha \gamma = (-\delta + \sum b_i \log_\alpha p_i)(\bmod n)$, $x$ can be obtained.

The index-calculus algorithm is suitable for the discrete logarithm problem over finite field, whose computation complexity is sub-exponential [10].

Shor's quantum algorithm can solve the discrete logarithm problem in polynomial time [2].

## 3 *t*-MDLP

***Definition 2***(*t*-MDLP) Suppose $N = p_1^{a_1} p_2^{a_2} \cdots p_c^{a_c}$, $g_1, g_2, \cdots, g_t \in Z_N$ and $\gcd(g_i, N) = 1$. The order of $g_i$ is $r_i$ and $g_i^{v_i} \notin \langle g_1, \cdots, g_{i-1}, g_{i+1}, \cdots, g_t \rangle$. $\langle g_1, g_2, \cdots, g_t \rangle$ is a group generated by $g_1, g_2, \cdots, g_t$, whose operation rule is modular multiplication. The t-multiple discrete logarithm of $\beta$ to the base $g_1, g_2, \cdots, g_t$ is the unique integer $k_1, k_2, \cdots, k_t$, such that $\beta = g_1^{k_1} g_2^{k_2} \cdots g_t^{k_t} \bmod N$. ($1 \leq i \leq t$、$1 \leq v_i \leq r_i - 1$ and $0 \leq k_i \leq r_i - 1$)

In fact, *t*-MDLP is the composition of discrete logarithm problem. Thus *t*-MDLP is harder than or equivalent to discrete logarithm problem under classical computation mode.

The property of *t*-MDLP is analyzed as below.

***Property 1*** *t*-MDLP can't degenerate into (*t*-1)-MDLP.

**Proof**: In definition 2, if *t*-MDLP degenerate into (*t*-1)-MDLP, there are $k_1', \cdots, k_{i-1}', k_{i+1}', \cdots, k_t'$ satisfying $\beta = g_1^{k_1'} \cdots g_{i-1}^{k_{i-1}'} g_{i+1}^{k_{i+1}'} \cdots g_t^{k_t'} \bmod N$.

Since

$$\beta = g_1^{k_1} g_2^{k_2} \cdots g_t^{k_t} \bmod N$$

we can obtain

$$g_1^{k_1} g_2^{k_2} \cdots g_t^{k_t} = g_1^{k_1'} \cdots g_{i-1}^{k_{i-1}'} g_{i+1}^{k_{i+1}'} \cdots g_t^{k_t'} \bmod N$$

i.e.

$$g_i^{k_i} = g_1^{k_1' - k_1} \cdots g_{i-1}^{k_{i-1}' - k_{i-1}} g_{i+1}^{k_{i+1}' - k_{i+1}} \cdots g_t^{k_t' - k_t} \bmod N$$

And it contradicts with $g_i^{v_i} \notin \langle g_1, \cdots, g_{i-1}, g_{i+1}, \cdots, g_t \rangle$.

Thus we can obtain this property.

∎

Property 1 shows that *t*-MDLP has non-degeneracy.

***Property 2*** The solution of *t*-MDLP is unique.

**Proof**: In definition 2, the solution of *t*-MDLP isn't unique, there are $k_1', k_2', \cdots, k_t'$ satisfying $\beta = g_1^{k_1'} g_2^{k_2'} \cdots g_t^{k_t'} \bmod N$. And $k_1', k_2', \cdots, k_t'$ is identical to $k_1, k_2, \cdots, k_t$. We can suppose $k_1 \neq k_1'$.

Since $\beta = g_1^{k_1} g_2^{k_2} \cdots g_t^{k_t} = g_1^{k_1'} g_2^{k_2'} \cdots g_t^{k_t'} \bmod N$, we can obtain

$$g_1^{k_1-k_1'} = g_2^{k_2'-k_2} g_3^{k_3'-k_3} \cdots g_t^{k_t'-k_t} \mod N.$$

And it contradicts with $g_1^{v_1} \notin \langle g_2, g_3, \cdots, g_t \rangle$.

Thus we can obtain this property.

∎

## 4 The solving difficulty of *t*-MDLP

### 1. The parameter selection affecting the solving difficulty of *t*-MDLP

Property 1 shows that *t*-MDLP can't degenerate into (*t*-1)-MDLP. However, it isn't clear whether special $k_1, k_2, \cdots, k_t$ will reduce the solving difficulty of *t*-MDLP. The analysis is shown as follow.

**Case 1**: How to avoid the solving difficulty of *t*-MDLP being equivalent to discrete logarithm problem

If there is $k$ satisfying

$$\begin{cases} k = k_1 \mod r_1 \\ k = k_2 \mod r_2 \\ \vdots \\ k = k_t \mod r_t \end{cases}$$

we can obtain

$$\beta = (g_1, g_2, \cdots, g_t)^k \mod N.$$

Thus *t*-MDLP is equivalent to discrete logarithm problem. And Shor's quantum algorithm can solve *t*-MDLP. How to avoid this case will be analyzed as below.

***Lemma 1***[11] The sufficient and necessary condition for the solvability of $\begin{cases} x = b_1 \mod m_1 \\ x = b_2 \mod m_2 \end{cases}$ is $\gcd(m_1, m_2) | (b_1, b_2)$. And the solvability is unique under module $\operatorname{lcm}(m_1, m_2)$. $\gcd(m_1, m_2)$ and $\operatorname{lcm}(m_1, m_2)$ are respectively the greatest common divisor and least common multiple of $m_1$ and $m_2$.

***Theorem 1*** Suppose $N = p_1^{a_1} p_2^{a_2} \cdots p_c^{a_c}$, $g_1, g_2, \cdots, g_t \in Z_N$ and $\gcd(g_i, N) = 1$. The order of $g_i$ is $r_i$ and $g_i^{v_i} \notin \langle g_1, \cdots, g_{i-1}, g_{i+1}, \cdots, g_t \rangle$. $\langle g_1, g_2, \cdots, g_t \rangle$ is a group generated by $g_1, g_2, \cdots, g_t$, whose operation rule is modular multiplication. Let $\beta = g_1^{k_1} g_2^{k_2} \cdots g_t^{k_t} \mod N \in \langle g_1, g_2, \cdots, g_t \rangle$. If $\gcd(r_{j_1}, r_{j_2}) \nmid (k_{j_1} - k_{j_2})$, there is no $k$ satisfying $\beta = (g_1, g_2, \cdots, g_t)^k \mod N$, where $1 \leq i \leq t$, $1 \leq v_i \leq r_i - 1$, $j_1, j_2 \in \{1, 2, \cdots, t\}$ and $j_1 \neq j_2$.

**Proof**: Since $g_i^{j_i} \notin \langle g_1, \cdots, g_{i-1}, g_{i+1}, \cdots, g_t \rangle$ for any $i$ and $j_i (1 \leq i \leq t, 1 \leq j_i \leq r_i - 1)$, to prove the theorem, we only need to prove that there is no solution for

$$\begin{cases} x = k_1 \mod r_1 \\ x = k_2 \mod r_2 \\ \vdots \\ x = k_t \mod r_t \end{cases} \quad (2)$$

If (2) has a solution, $\begin{cases} x = k_i \mod r_i \\ x = k_j \mod r_j \end{cases}$ also has a solution, where $i, j \in \{1, 2, \cdots, t\}$.

Since there are $j_1, j_2 \in \{1, 2, \cdots, t\}$ satisfying $j_1 \neq j_2$ and $\gcd(r_{j_1}, r_{j_2}) \nmid (k_{j_1} - k_{j_2})$, $\begin{cases} x = k_{j_1} \mod r_{j_1} \\ x = k_{j_2} \mod r_{j_2} \end{cases}$ has no solution.

Thus we obtain a contradiction and the theorem is correct.

∎

***Corollary 1*** In theorem 1, if $k$ satisfy $\beta = (g_1 g_2 \cdots g_t)^k \mod N$, $k$ is unique under module

$lcm(r_1, r_2, \cdots, r_t)$.

**Proof**: According to lemma 1, we can obtain corollary 1.

∎

According to theorem 1, if there is $j_1, j_2 \in \{1, 2, \cdots, t\}$ which satisfy $j_1 \neq j_2$ and $\gcd(r_{j_1}, r_{j_2}) \nmid (k_{j_1} - k_{j_2})$, it's certain that the solving difficulty of *t*-MDLP isn't equivalent to discrete logarithm problem, i.e. there is no $k$ satisfying $\beta = g_1^{k_1} g_2^{k_2} \cdots g_t^{k_t} = g_1^k g_2^k \cdots g_t^k \mod N$.

According to corollary 1, the number of $k$ is $lcm(r_1, r_2, \cdots, r_t)$, which satisfy $\beta = g_1^{k_1} g_2^{k_2} \cdots g_t^{k_t} = g_1^k g_2^k \cdots g_t^k \mod N$.

**Case 2**: How to avoid the solving difficulty of *t*-MDLP being equivalent to one discrete logarithm problem and one (*t*-1)-MDLP

Since $\beta = g_1^{k_1} g_2^{k_2} \cdots g_t^{k_t} \mod N$, if there is $i$ satisfying $g_2^{k_2} g_3^{k_3} \cdots g_t^{k_t} = 1 \mod p_j$ (let $i = 1$),

$$\beta = g_i^{k_i} \mod p_j.$$

And the solving difficulty of $\beta = g_1^{k_1} g_2^{k_2} \cdots g_t^{k_t} \mod N$ is equivalent to $\beta = g_i^{k_i} \mod p_j$ and $\beta g_1^{-k_1} = g_2^{k_2} \cdots g_t^{k_t} \mod N$.

Furthermore, if there is $i'$ satisfying $g_3^{k_3} \cdots g_t^{k_t} = 1 \mod p_{i'}$ (let $i' = 2$),

$$\beta g_1^{-k_1} = g_2^{k_2} \mod p_{i'}$$

And the solving difficulty of $\beta = g_1^{k_1} g_2^{k_2} \cdots g_t^{k_t} \mod N$ is equivalent to two discrete logarithm problems and $\beta g_1^{-k_1} g_2^{-k_2} = g_3^{k_3} \cdots g_t^{k_t} \mod N$.

Similarly, the solving difficulty of *t*-MDLP is equivalent to $k$ discrete logarithm problems and one (*t-k*)-MDLP. Thus the difficulty will be reduced based on quantum algorithm. How to avoid the case will be analyzed as follows.

**Theorem 2** *Suppose* $N = p_1^{a_1} p_2^{a_2} \cdots p_c^{a_c}$, $g_1, g_2, \cdots, g_t \in Z_N$ *and* $\gcd(g_i, N) = 1$. *The order of* $g_i$ *is* $r_i$ *and* $g_i^{v_i} \notin \langle g_1, \cdots, g_{i-1}, g_{i+1}, \cdots, g_t \rangle$. $\langle g_1, g_2, \cdots, g_t \rangle$ *is a group generated by* $g_1, g_2, \cdots, g_t$, *whose operation rule is modular multiplication. Let* $\beta = g_1^{k_1} g_2^{k_2} \cdots g_t^{k_t} \mod N \in \langle g_1, g_2, \cdots, g_t \rangle$. *If there is no* $i$ *satisfying* $g_1^{k_1} \cdots g_{i-1}^{k_{i-1}} g_{i+1}^{k_{i+1}} \cdots g_t^{k_t} \neq 1 \mod p_j$, $\beta \neq g_i^{k_i} \mod p_j$, *where* $1 \leq i \leq t$, $1 \leq v_i \leq r_i - 1$, *and* $1 \leq j \leq c$.

**Proof**: According to $\gcd(g_i, N) = 1$, if $\beta = g_i^{k_i} \mod p_j$, there is $x$ satisfying

$$\beta x = x g_i^{k_i} = 1 \mod p_j$$

Since $\beta = g_1^{k_1} g_2^{k_2} \cdots g_t^{k_t} \mod N$,

$$\beta = g_1^{k_1} g_2^{k_2} \cdots g_t^{k_t} \mod p_j$$

Furthermore, we can obtain that

$$\beta x = x g_1^{k_1} g_2^{k_2} \cdots g_t^{k_t} \mod p_j$$

i.e.

$$g_1^{k_1} \cdots g_{i-1}^{k_{i-1}} g_{i+1}^{k_{i+1}} \cdots g_t^{k_t} = 1 \mod p_j$$

It is in contradiction with $g_1^{k_1} \cdots g_{i-1}^{k_{i-1}} g_{i+1}^{k_{i+1}} \cdots g_t^{k_t} \neq 1 \mod p_j$.

Thus we can obtain this theorem.

∎

According to theorem 2, if there is no $i$ satisfying $g_1^{k_1} \cdots g_{i-1}^{k_{i-1}} g_{i+1}^{k_{i+1}} \cdots g_t^{k_t} \neq 1 \mod p_j$, it's certain that the solving difficulty of *t*-MDLP isn't equivalent to one discrete logarithm problem and one (*t*-1)-MDLP.

According to case 1 and case 2, if $\gcd(r_{j_1}, r_{j_2}) \nmid (k_{j_1} - k_{j_2})$, and there is no $i$ satisfying

$g_1^{k_1} \cdots g_{i-1}^{k_{i-1}} g_{i+1}^{k_{i+1}} \cdots g_t^{k_t} \neq 1 \bmod p_j$, the solving difficulty of $t$-MDLP will not reduce, where $j_1, j_2 \in \{1,2,\cdots,t\}$ and $j_1 \neq j_2$.

The existence of $t$-MDLP will be illustrated as follow.

If $N=35$, $t=2$, $g_1=13$ and $g_2=19$, $r_1=4$ and $r_2=4$. And the truth table of $g_1^{k_1} g_2^{k_2} \bmod N$ is as Table.1.

Table.1 the truth table of $g_1^{k_1} g_2^{k_2} \bmod N$

| $k_2$ \ $k_1$ | 1 | 2 | 3 | 4 |
|---|---|---|---|---|
| 1 | 2 | 26 | **23** | 19 |
| 2 | 3 | 4 | 17 | 11 |
| 3 | **22** | 6 | 8 | 34 |
| 4 | 13 | 29 | 27 | 1 |

According to Table.1, when $(k_1,k_2)$ is $(3,1)$ or $(1,3)$, $\gcd(r_1,r_2) \nmid (k_1-k_2)$, $g_1^3 \neq 1 \bmod 5$, $g_1^3 \neq 1 \bmod 7$, $g_1 \neq 1 \bmod 5$, $g_1 \neq 1 \bmod 7$, $g_2^3 \neq 1 \bmod 5$, $g_2^3 \neq 1 \bmod 7$, $g_2 \neq 1 \bmod 5$ and $g_2 \neq 1 \bmod 7$. Thus the solving difficulty of $t$-MDLP will not reduce when $(k_1,k_2)$ is $(3,1)$ or $(1,3)$.

## 2. The solving difficulty of $t$-MDLP under classical computing mode

**Exhaustive method**

Since $g_i^{v_i} \notin \langle g_1,\cdots,g_{i-1},g_{i+1},\cdots,g_t \rangle$, $k_i$ can't be quickly solved from $k_1,\cdots,k_{i-1},k_{i+1},\cdots,k_t$ and $g_1,\cdots,g_{i-1},g_{i+1},\cdots,g_t$ for $i=1,2,\cdots,t$, i.e. $k_1,k_2,\cdots,k_t$ can't be obtained quickly.

Although there is no $i$ satisfying $g_1^{k_1} \cdots g_{i-1}^{k_{i-1}} g_{i+1}^{k_{i+1}} \cdots g_t^{k_t} \neq 1 \bmod p_j$ for arbitrary $p_j$, we can't quickly obtain $k_1,k_2,\cdots,k_t$, i.e. the computation complexity of exhaustive methods can't be reduced. For each $k \in \{1,2,\cdots,lcm(r_1,r_2,\cdots,r_t)\}$, there are $k_1,k_2,\cdots,k_t$ satisfying $\begin{cases} k = k_1 \bmod r_1 \\ k = k_2 \bmod r_2 \\ \vdots \\ k = k_t \bmod r_t \end{cases}$. And $k_1,k_2,\cdots,k_t$ can't be assumed these special values in practical use. Thus the computation complexity of exhaustive method is $r_1 r_2 \cdots r_t - lcm(r_1,r_2,\cdots,r_t)$.

In definition 2, the order of $g_1,g_2,\cdots,g_t$ is unknown to attackers. Thus, the computation complexity of $t$-MDLP will not be reduced by their order.

**Index-calculus algorithm**

The index-calculus algorithm[10] is the most powerful method known for computing discrete logarithm. The technique employed does not apply to all groups, such as cyclic group $G$ of order $n$, but when it does, it often gives a subexponential-time algorithm. First, the algorithm constructs the linear equations of $\log_\alpha p_1 (\bmod n), \log_\alpha p_2 (\bmod n), \cdots, \log_\alpha p_m (\bmod n)$. Then it can obtain a system of linear equations. Furthermore, $\log_\alpha p_1 (\bmod n), \log_\alpha p_2 (\bmod n), \cdots, \log_\alpha p_m (\bmod n)$ can be obtained. Finally, $\gamma \alpha^\delta$ is tried to write as a product of elements in $S=\{p_1,p_2,\cdots,p_m\}$: $\gamma \alpha^\delta = \sum_{i=1}^{m} p_i^{b_i}$. And $\log_\alpha \gamma$ can be obtained by $\log_\alpha \gamma = (-\delta + \sum b_i \log_\alpha p_i)(\bmod n)$.

If the algorithm is applied to solve $t$-MDLP, we can obtain $\log_\alpha \beta (\bmod r)$, $\log_\alpha g_1 (\bmod r)$, $\log_\alpha g_2 (\bmod r), \ldots, \log_\alpha g_t (\bmod r)$, where $r$ is the order of $\alpha$ under modulo $N$.

Since $\beta = g_1^{k_1} g_2^{k_2} \cdots g_t^{k_t} \bmod N$,

$$\log_\alpha \beta = (\sum_{i=1}^{t} k_i \log_\alpha g_i)(\bmod r) \qquad (3)$$

For another $\alpha'$, we can obtain

$$\log_{\alpha'} \beta = (\sum_{i=1}^{t} k_i \log_{\alpha'} g_i)(\bmod r') \qquad (4)$$

where $r'$ is the order of $\alpha'$ under modulo $N$.

If $r \neq r'$, the compute mode of (3) is different from (4). Thus $k_1, k_2, \cdots, k_t$ can't be solved by constructing the system of linear equations.

If $r = r'$, since $\log_{\alpha'} x = \dfrac{\log_\alpha x}{\log_\alpha \alpha'}$,

$$\log_{\alpha'} \beta = \frac{\log_\alpha \beta}{\log_\alpha \alpha'} = (\frac{1}{\log_\alpha \alpha'} \sum_{i=1}^{t} k_i \log_\alpha g_i)(\bmod r)$$

Thus (3) is equivalent to (4), i.e. $k_1, k_2, \cdots, k_t$ can't be solved by constructing the system of linear equations.

In conclusion, the index-calculus algorithm is not suit for solving $t$-MDLP.

### 3. The solving difficulty of $t$-MDLP under quantum computing mode

At present, most of the quantum algorithm with polynomial time can be reduced to the quantum algorithm for the hidden subgroup problem, which is a general algorithm.

If the parameter of $t$-MDLP doesn't satisfy the case in theorem 1, i.e. there is $k$ satisfying $\beta = g_1^{k_1} g_2^{k_2} \cdots g_t^{k_t} = (g_1 g_2 \cdots g_t)^k \bmod N$, $k$ can be obtained by Shor's algorithm.

Since $g_i^{v_i} \notin \langle g_1, \cdots, g_{i-1}, g_{i+1}, \cdots, g_t \rangle$,

$$\begin{cases} k_1 = k \bmod r_1 \\ k_2 = k \bmod r_2 \\ \vdots \\ k_t = k \bmod r_t \end{cases}$$

i.e. Shor's algorithm can be applied to solve $t$-MDLP

If the parameter of $t$-MDLP doesn't satisfy the case in theorem 2, $\beta = g_i^{k_i} \bmod p_j$. And $k_i$ can be obtained by Shor's algorithm. And $k_1, \cdots, k_{i-1}, k_{i+1}, \cdots, k_t$ satisfy $1 = g_1^{k_1} \cdots g_{i-1}^{k_{i-1}} g_{i+1}^{k_{i+1}} \cdots g_t^{k_t} \bmod N$. In other words, the computation complexity of $t$-MDLP will be reduced.

Thus, Shor's algorithm is suit for solving $t$-MDLP when the parameter doesn't satisfy the cases in theorem 1 and 2. And we can obtain theorem 3.

**Theorem 3** Suppose $N = p_1^{a_1} p_2^{a_2} \cdots p_c^{a_c}$, $g_1, g_2, \cdots, g_t \in Z_N$ and $\gcd(g_i, N) = 1$. The order of $g_i$ is $r_i$ and $g_i^{v_i} \notin \langle g_1, \cdots, g_{i-1}, g_{i+1}, \cdots, g_t \rangle$. $\langle g_1, g_2, \cdots, g_t \rangle$ is a group generated by $g_1, g_2, \cdots, g_t$, whose operation rule is modular multiplication. Let $\beta = g_1^{k_1} g_2^{k_2} \cdots g_t^{k_t} \bmod N \in \langle g_1, g_2, \cdots, g_t \rangle$. There are two necessary conditions of which $t$-MDLP resists Shor's quantum algorithm: $\gcd(r_{j_1}, r_{j_2}) \nmid (k_{j_1} - k_{j_2})$ and there is no $i$ satisfying $g_1^{k_1} \cdots g_{i-1}^{k_{i-1}} g_{i+1}^{k_{i+1}} \cdots g_t^{k_t} \neq 1 \bmod p_j$. ($1 \leq i \leq t$, $1 \leq v_i \leq r_i - 1$, $j_1, j_2 \in \{1, 2, \cdots, t\}$, $j_1 \neq j_2$, $1 \leq v_i \leq r_i - 1$, and $1 \leq j \leq c$)

**Proof**: According to theorem 1 and 2, we can obtain theorem 3.

∎

As we know, there is a polynomial-time quantum algorithm for the hidden subgroup problem [4]. If the problem can be reduced to the hidden subgroup problem, it can be solved, such as factorization and discrete logarithm problem. And Shor's quantum algorithm is a particular circumstance of the quantum algorithm for the hidden subgroup problem [4]. If the parameters of *t*-MDLP don't satisfy two conditions in Theorem 3, Shor's quantum algorithm will reduce the solving difficulty, i.e the quantum algorithm for the hidden subgroup problem can also do. Thus the two conditions in Theorem 3 are also the necessary conditions of which *t*-MDLP resists the quantum algorithm for the hidden subgroup.

## 5 Conclusion

In this paper, *t*-MDLP is presented, whose solving difficulty is analyzed. Ant two necessary conditions of resisting the quantum algorithm for the hidden subgroup are given. Based on the problem, designing the public-key crypto is worth further studying.

## Acknowledgements

The authors gratefully acknowledge the financial support from the National Basic Research Program of China (Grant No. 2013CB338002) and the National Natural Science Foundation of China (Grant No. 61502526).